%% file: main_rev.tex
\documentclass[sigconf]{acmart}

\AtBeginDocument{%
  }

\copyrightyear{2026}
\acmYear{2026}
\setcopyright{cc}
\setcctype{by}
\acmConference[ICSE '26]{2026 IEEE/ACM 48th International Conference on Software Engineering}{April 12--18, 2026}{Rio de Janeiro, Brazil}
\acmBooktitle{2026 IEEE/ACM 48th International Conference on Software Engineering (ICSE '26), April 12--18, 2026, Rio de Janeiro, Brazil}
\acmPrice{}
\acmDOI{10.1145/3744916.3787768}
\acmISBN{979-8-4007-2025-3/2026/04}

\usepackage{graphicx}
\usepackage{hyperref}
\usepackage{textcomp}
\usepackage{forest}
\usepackage{subcaption}
\usepackage{listings}
\usepackage{tabularx}
\usepackage{threeparttable}
\usepackage{booktabs}
\usepackage{makecell}
\usepackage{adjustbox}
\usepackage{rotating}
\usepackage{array}
\usepackage{caption}
\usepackage{url}
\usepackage{tikz}
\usepackage{tcolorbox}
\usepackage{enumitem}
\setlist[enumerate]{leftmargin=1.5em}
\setlist[itemize]{leftmargin=1.5em}
\usepackage{array}
\usepackage{soul}
\usepackage{etoolbox}
\usepackage{algorithm}
\usepackage{multirow}
\usepackage{seqsplit} %
\usepackage[noend]{algpseudocode}
\usepackage[dvipsnames]{xcolor}

\input{text_rev/macro.tex}

\usetikzlibrary{shapes.geometric, arrows}
\tikzstyle{startstop} = [rectangle, rounded corners, minimum width=3cm, minimum height=1cm, text centered, draw=black, fill=red!30]
\tikzstyle{process} = [rectangle, minimum width=3cm, minimum height=1cm, text centered, draw=black, fill=orange!30]
\tikzstyle{decision} = [diamond, minimum width=3cm, minimum height=1cm, text centered, draw=black, fill=green!30]
\tikzstyle{arrow} = [thick,->,>=stealth]

\begin{document}

\title[\methodNameLong{}: Last-Mile Regression Test Augmentation]{\methodNameLong{}:\\ Last-Mile, Pull Request-Based Regression Test Augmentation}

\author{Zitong Zhou}
\authornote{Both authors contributed equally to this work}
\affiliation{%
  \institution{UCLA}
  \city{Los Angeles}
  \country{USA}
}
\email{zitongzhou@cs.ucla.edu}

\author{Matteo Paltenghi}
\authornotemark[1]
\affiliation{%
  \institution{University of Stuttgart}
  \city{Stuttgart}
  \country{Germany}
}
\email{mattepalte@live.it}

\author{Miryung Kim}
\affiliation{%
  \institution{UCLA}
  \city{Los Angeles}
  \country{USA}
}
\email{miryung@cs.ucla.edu}

\author{Michael Pradel}
\affiliation{%
  \institution{CISPA Helmholtz Center for Information Security}
  \city{Stuttgart}
  \country{Germany}
}
\email{michael@binaervarianz.de}

\renewcommand{\shortauthors}{Zitong Zhou, Matteo Paltenghi, Miryung Kim, and Michael Pradel}

\begin{abstract}
Software is in constant evolution, with developers frequently submitting pull requests (PRs) to introduce new features or fix bugs.
Testing newly added or modified code in PRs is critical to maintaining software quality.
Yet, even in projects with extensive test suites, some of the code modified in PRs may remain untested, leaving a ``last-mile'' regression test gap.
Existing automated test generators mostly focus on improving overall code coverage, but do not specifically target the uncovered lines in PRs.
This paper presents \methodNameLong{} (\methodName{}), a novel, LLM-based test augmentation technique that specifically addresses the last-mile regression test gap in PRs.
Our approach is enabled by three key contributions:
(i) Instead of focusing on overall code coverage, \methodName{} considers a specific PR and the lines left uncovered after applying the PR, offering developers augmented tests for code just when it is on the developers' mind.
(ii) We identify providing suitable test context as a crucial challenge for an LLM to generate useful tests, and present two techniques to extract relevant test content, such as existing test functions, fixtures, and data generators.
(iii) To make augmented tests acceptable for developers, \methodName{} carefully integrates them into the existing test suite, e.g., by matching the test's structure and style with the existing tests, and generates a summary of the test addition for developer review.
We evaluate \methodName{} on 145 PRs from three popular, complex, and well-tested open-source projects\textemdash SciPy, Qiskit, and Pandas.
The approach successfully helps \ratioPRFullCov{} of PRs achieve \textit{full patch coverage}, at the affordable cost of \avgCostPerPR{} per PR, demonstrating its effectiveness and feasibility.
A qualitative assessment of the generated tests shows that human reviewers find the tests to be worth adding (\avgScoreWorthwhile{}/5.0), well integrated (\avgScoreWellIntegrated{}/5.0), and relevant to the PR (\avgScoreRelatedToPr{}/5.0).
Ablation studies show test context is crucial for context-aware test generation, leading to \avgCovImprvOverNoTC{} coverage.
In a contribution study, we submitted \totalSubmittedTests{} tests to these projects, of which \totalMergedTests{} have already been merged, and two previously unknown bugs were discovered and fixed.
We envision our approach to be integrated into CI workflows, automating the last mile of regression test augmentation.

\end{abstract}


\settopmatter{printacmref=true,printfolios=true}

\maketitle

\lstdefinestyle{pythonstyle}{
  language=Python,
  basicstyle=\small\ttfamily,
  keywordstyle=\color{blue},
  commentstyle=\color{green!60!black},
  stringstyle=\color{red},
  showstringspaces=false,
  breaklines=true,
  postbreak=\mbox{\textcolor{red}{$\hookrightarrow$}\space},
  tabsize=4,
  frame=single,
  numbers=left,
  numberstyle=\tiny\color{gray}
}

\input{text_rev/intro.tex}

\input{text_rev/problem.tex}

\input{text_rev/approach.tex}

\input{text_rev/implementation.tex}
\input{text_rev/eval.tex}

\input{text_rev/threats.tex}

\input{text_rev/related_work.tex}

\section{Conclusion}
We presented \methodName{}, a novel approach for augmenting regression tests at the PR-level by targeting the ``last-mile'' gap in patch coverage.
\methodName{} leverages LLMs, enriched with PR and test context, to generate and integrate meaningful tests that align with project-specific conventions.
Our evaluation across 145 PRs from SciPy, Qiskit, and Pandas proves \methodName{}'s effectiveness, achieving full patch coverage in 30\% of cases, discovering previously unknown bugs, and receiving positive developer feedback.
With low per-PR cost and high acceptability, \methodName{} offers a practical path toward automating fine-grained test augmentation in CI workflows.

\begin{acks}
This work is supported by the National Science Foundation under grant numbers 2426162, 2106838, and 2106404, by the European Research Council (ERC; grant agreements 851895 and 101155832), and by the German Research Foundation (DFG; projects 492507603, 516334526, and 526259073). It is also supported in part by funding from Amazon and Samsung. We thank the reviewers for their constructive feedback that helped improve the work. 
\end{acks}

\bibliographystyle{ACM-Reference-Format}
\bibliography{phd-mattepalte,referencesMichael,newreference}
\end{document}

%% file: text_rev/macro.tex
\newcommand{\methodNameLong}{Change And Cover}
\newcommand{\methodName}{ChaCo}

\newtcolorbox{resultbox}{
    colback=black!5!white,
    colframe=white!50!black,
    boxsep=0mm,
}

\newcommand{\code}[1]{\lstinline[breaklines=true,basicstyle=\ttfamily]{#1}}

\lstdefinestyle{mypythonstyle}{
    frame=tb,
    language=Python,
    basicstyle=\color{black!}\scriptsize\ttfamily, %
    commentstyle=\color{green!50!black}\ttfamily, %
    breaklines=true, %
    tabsize=2, %
    morecomment=[l]{//}, %
    morecomment=[s]{/*}{*/}, %
    stringstyle=\color{red!80!black}, 
    emph={%
        STRING%
        },emphstyle={\color{blue}\ttfamily},%
    alsoletter={,:,|,;,+,?,*,\(,\)}, %
    keywords={def, return, if, else, elif, while, for, in, with, as, import, from, try, except, class, lambda, pass, break, continue, True, False, None, and, or},
    keywordstyle={\color{blue}},
    moredelim=[is][\textcolor{magenta}]{|}{|},
    moredelim=[is][\textcolor{orange}]{^}{^},
}

\makeatletter
\newcommand{\highlightline}[1]{%
  \begingroup
  \setlength{\fboxsep}{0pt}%
  \colorbox{gray!15}{\strut #1\hspace*{\fill}}%
  \endgroup
}
\makeatother
\newcommand{\highlightlineRed}[1]{%
  \begingroup
  \setlength{\fboxsep}{0pt}%
  \colorbox{red!25}{\strut #1\hspace*{\fill}}%
  \endgroup
}
\newcommand{\highlightlineOrange}[1]{%
  \begingroup
  \setlength{\fboxsep}{0pt}%
  \colorbox{yellow!70}{\strut #1\hspace*{\fill}}%
  \endgroup
}

\lstdefinestyle{mypythonstyleDiff}{
  style=mypythonstyle,
  literate={+}{{\textcolor{ForestGreen}{+}}}{1},
  moredelim=[is][\highlightline]{@HL@}{@endHL@},
  moredelim=[is][\highlightlineRed]{@RED@}{@endRED@},
  moredelim=[is][\highlightlineOrange]{@YELLOW@}{@endYELLOW@}
}

\newcommand{\totalSubmittedTests}{12}
\newcommand{\totalMergedTests}{8}
\newcommand{\totalUnderReviewTests}{4}

\newcommand{\avgCostPerPR}{\$0.11}

\newcommand{\totalCovLinesAdded}{189}
\newcommand{\totalPRFullCov}{43}
\newcommand{\ratioPRFullCov}{30\%}

\newcommand{\avgPassRateImprvOverNoFBPercent}{312\%}

\newcommand{\avgCovImprvOverLLMPercent}{100\%}
\newcommand{\avgCovImprvOverNoTC}{2$\times$}

\newcommand{\avgCovImprvOverNoFB}{5.6$\times$}
\newcommand{\avgCovImprvOverNoFBPercent}{460\%}

\newcommand{\avgScoreWorthwhile}{4.53}
\newcommand{\avgScoreWellIntegrated}{4.20}
\newcommand{\avgScoreRelatedToPr}{4.70}
\newcommand{\percAgrWorthwhile}{60.00\%}
\newcommand{\percAgrWellIntegrated}{66.67\%}
\newcommand{\percAgrRelatedToPr}{66.67\%}

%% file: text_rev/intro.tex
\section{Introduction}

Testing newly added or modified code in pull requests (PRs) is critical to maintaining software quality. For example, SciPy, a popular scientific computing library, claims that it ``aims for a high coverage for all new code that is added''~\cite{manualscipy}. Tools like Codecov \cite{codecov} that measure code coverage as projects evolve are used in Continuous Integration (CI) by projects, such as NodeJS, Keras, and scikit-learn~\cite{nodejs,keras,scikitlearn}.
We call the coverage of code changed in a PR \emph{patch coverage}.
While many developers already add tests to exercise changed code, we observe that often a few lines of uncovered code remain, which we refer to as the \emph{last-mile regression test gap}.

There is an extensive body of prior work on test generation and fuzzing, yet few techniques sufficiently address the last-mile problem in PRs.
One line of work focuses on improving the overall code coverage of a project~\cite{pizzornoCoverUpEffectiveHigh2025,Ryan2024,Lemieux2023,schaferEmpiricalEvaluationUsing2024,wangHITSHighcoverageLLMbased2024a,lukasczykPynguinAutomatedUnit2022}, but does not focus on PRs.
Another line of work targets code changes, e.g., by fuzzing newly/frequently changed code~\cite{zhangDeltaFuzzHistoricalVersion2022,zhuRegressionGreyboxFuzzing2021}.
However, these approaches do not consider the context of a PR, such as the discussion between developers, the PR description, and links to related issues or PRs.
Finally, Testora~\cite{icse2026-Testora} tries to detect unintended changes introduced by a PR, but it does not aim to improve patch coverage.

Despite advances in automated test generation, we identify several key research gaps: (1) Prior work has not focused on generating regression tests specifically to improve patch coverage in PRs; (2) Existing approaches rarely leverage the rich context available in PRs, such as links to related PRs and issues; (3) Test generation approaches often ignore the structure and utilities of the existing test suite, including fixtures, markers, mocks, and custom assertions.

We address these gaps with \methodNameLong{} (\methodName{}), a novel, LLM-based test augmentation technique that specifically addresses the last-mile regression test gap in PRs.
To address gaps (1) to (3), our approach
(1) considers missing patch coverage of a specific PR, offering developers augmented tests for code just when it is on their mind;
(2) provides suitable test context to the LLM, such as existing test functions, fixtures, and data generators;
(3) carefully integrates tests into the existing test suite, e.g., by matching the test's structure and style with the existing tests.
Furthermore, we evaluate \methodName{} not only on automated metrics, such as coverage improvement, but also by manual review and real-world acceptability.

PRs are a natural playground for adding new tests because each PR represents a discrete unit of reviewable changes to a codebase. A PR typically introduces new functionality, bug fixes, or performs a refactoring, making it an ideal opportunity to ensure that the affected code is adequately tested before integration. By focusing on PRs, we can catch gaps in test coverage and prevent regressions early. Moreover, PRs typically provide rich contextual information, including change descriptions, related issues, and developer discussions, which can be leveraged to generate targeted and relevant tests.

\begin{figure}[t]
  \centering
\begin{subfigure}[b]{0.48\textwidth}
    \centering
    \includegraphics[width=0.80\textwidth]{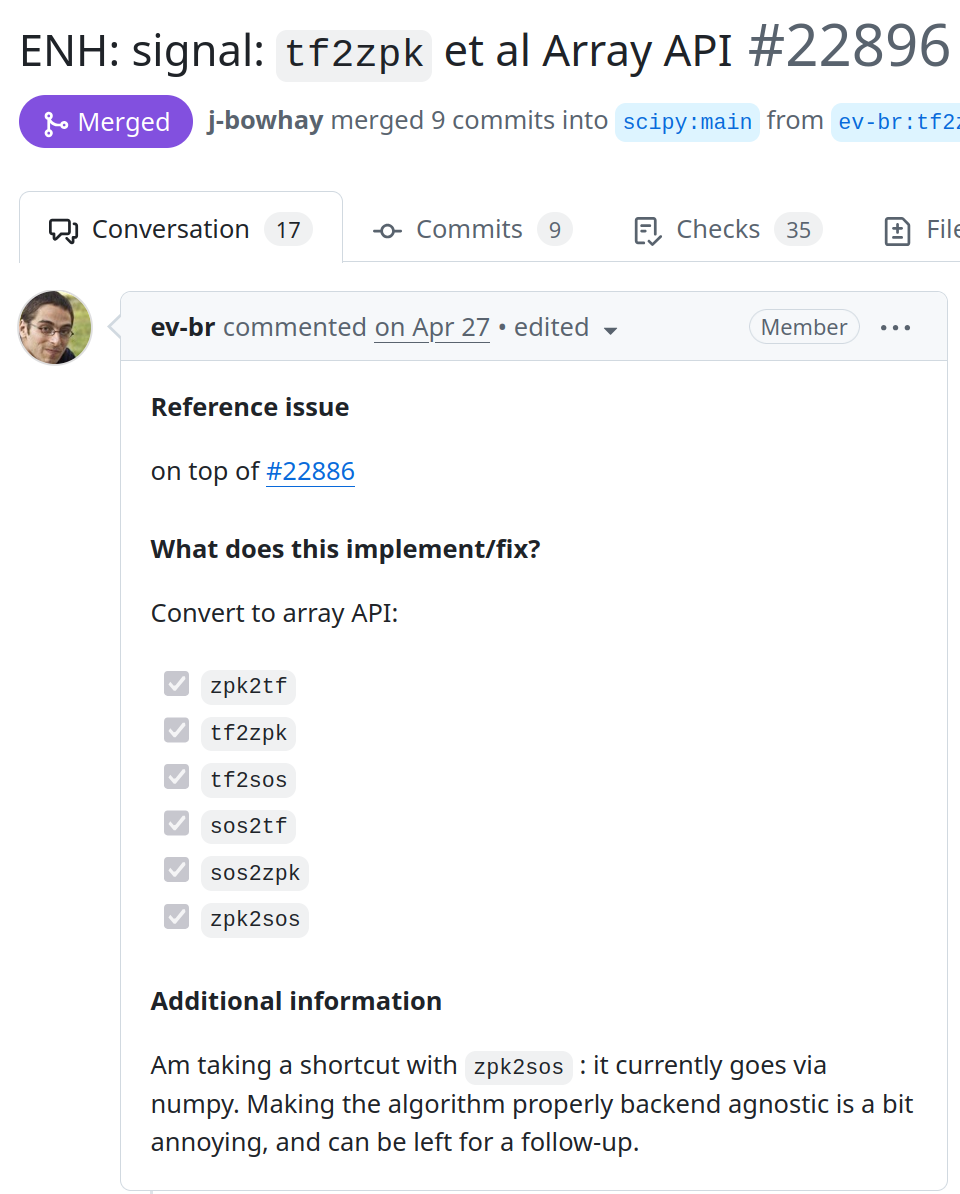}
    \caption{PR with title, description, discussion, and links.}
    \label{fig:input_pr}
\end{subfigure}

\vspace{3mm}
\begin{subfigure}[b]{0.38\textwidth}
  \centering
\begin{lstlisting}[
      style=mypythonstyleDiff
    ]
def zpk2tf(z, p, k):
    """
    Return polynomial transfer function ...
    """
+   xp = array_namespace(z, p)
+   z, p, k = map(xp.asarray, (z, p, k))
+   ...
+   @YELLOW@if z.ndim > 1:@endYELLOW@
+       @RED@temp = _pu.poly(z[0], xp=xp)@endRED@
+       @RED@b = xp.empty((z.shape...),...)@endRED@
+       @RED@if k.shape[0] == 1:@endRED@
            k = [k[0]] * z.shape[0]
        for i in range(z.shape[0]):
+           @RED@b[i] = k[i] * _pu.poly(z[i], xp=xp)@endRED@
    else:
+       b = k * _pu.poly(z, xp=xp)
\end{lstlisting}
    \caption{PR diff with missing patch coverage highlighted.}
    \label{fig:pr_code_example}
\end{subfigure}
\vspace{-.3em}

  \caption{Illustration of a PR's patch coverage gap. \methodName{} runs the regression test suite on the PR to identify the missing patch coverage: missing branches in yellow, lines in red.}
  \label{fig:motivating_scipy_pr}
  \vspace{-.6em}
\end{figure}

Figure~\ref{fig:input_pr} shows a SciPy PR as a motivating example.
The code changes in Figure~\ref{fig:pr_code_example} show the lines added by the PR, and the highlighted lines are the intersection between the changes, and those not covered by the test suite. \methodName{} contributed a test to this PR, which exposed a bug in the implementation that was subsequently fixed.
To further motivate the need for suitable \textit{test context}, consider a requirement for new tests added to SciPy: If a feature under test does not support GPUs, its test functions must be marked with the test marker \code{@pytest.mark.skip_xp_backends(cpu_only=True)}.
Without understanding this specific requirement, LLMs will likely fail to generate a test that developers would accept.
As another example, consider an existing test function \code{test\_foo} that already tests the function \code{foo}, to which the PR adds a new optional parameter.
When prompted to produce a test for the new parameter, it is best for the LLM to reference \code{test\_foo}, and generate a new test that matches the structure and style of \code{test\_foo}.

Our evaluation applies \methodName{} to generate tests for 145 recent real-world PRs from three open-source projects -- SciPy, Qiskit, and Pandas.
Our results show that \methodName{} successfully helps \ratioPRFullCov{} of PRs achieve \textit{full patch coverage}, at an average cost of only \avgCostPerPR{} per PR.
A qualitative assessment of the generated tests shows that human reviewers find the tests to be worth adding (\avgScoreWorthwhile{}/5.0), well integrated (\avgScoreWellIntegrated{}/5.0), and relevant to the PR (\avgScoreRelatedToPr{}/5.0).
We also conduct a contribution study that submitted \totalSubmittedTests{} of \methodName{}'s generated tests to these projects, of which \totalMergedTests{} have been merged and \totalUnderReviewTests{} are under review, demonstrating the practical utility of \methodName{} in real-world development. \methodName{}'s added tests have already exposed two bugs in SciPy, both confirmed and fixed.
Finally, our ablation study confirms that test context is crucial for generating high-quality tests: Compared to a variant of the approach without test context and without runtime feedback, \methodName{} achieves a \avgCovImprvOverNoTC{} and \avgCovImprvOverNoFB{} higher total coverage increment, respectively.

In summary, our work makes the following contributions:
\begin{itemize}[leftmargin=*]
\item \textbf{PR-specific test generation.} We introduce a novel approach that leverages PR context to generate regression tests specifically aimed at improving patch coverage.
\item \textbf{Context-aware test generation.} We integrate dynamic analysis with LLM-based techniques to leverage test context, ensuring tests align with existing testing practices.
\item \textbf{Test suite integration.} Our approach generates tests that are consistent with the structure and utilities of the existing test suite, increasing the likelihood of acceptance by maintainers.
\item \textbf{Real-world evaluation.} We evaluate our approach by augmenting real-world PRs from large open-source software with additional tests and assessing their acceptance and feedback from developers, demonstrating the practical utility of our method.
\item \textbf{Data availability.} \methodName{} is available at \url{https://github.com/UCLA-SEAL/Change-Cover}.
\end{itemize}

%% file: text_rev/problem.tex
\section{Problem Statement}

The overall goal of \methodName{} is to augment a given PR with additional test cases that cover otherwise uncovered code that was newly introduced or modified code by the PR. \methodName{} assumes the PR is bug-free and generates tests to catch future regressions. Exposing buggy code changes on the spot is an orthogonal problem that requires a different approach, such as Testora~\cite{icse2026-Testora}. Isolating these two goals is deliberate: bug detection hinges on a change-intent oracle~\cite{icse2026-Testora}, which is not required for \methodName{}.

\methodName{} takes a PR as input, which consists of a title, a description, a sequence of discussion comments, and a diff.
The diff consists of line-level additions, deletions, or modifications, as well as file-level additions or removals.
Among all lines touched by the diff, we focus on the set $\mathcal{E}$ of \emph{executable lines}, i.e., all non-comment, non-empty source code lines that are not in test files but in the main source code of the project.
We partition $\mathcal{E}$ into two subsets: the set $\mathcal{C}$ of lines that are covered by the existing tests and the set $\mathcal{U}$ of lines that remain uncovered.
Given these sets, we define the \emph{patch coverage} of the PR as the fraction of modified lines covered by existing test:
$\mathit{pc} = \frac{\lvert \mathcal{C} \rvert}{\lvert \mathcal{E} \rvert}$.
If a PR has a patch coverage of 100\%, we call it \emph{fully covered}. 
Based on the above definitions, the problem addressed by \methodName{} consists of two subproblems, \emph{PR-based test generation} and \emph{test integration}, as detailed in the following.

\subsection{Task 1: PR-Based Test Generation}
The first subtask is to increase the patch coverage of PRs that are not yet fully covered:
\begin{definition}[PR-based test generation task]
  Given a PR with patch coverage $\mathit{pc} < 100\%$, generate a test case $t$ such that:
  \begin{itemize}[leftmargin=*]
    \item $t$ covers at least one line in $\mathcal{U}$.
    \item $t$ is relevant to the changes introduced in the PR.
  \end{itemize}
\end{definition}

Meaningful tests should reuse project-specific or module-specific test utilities, including but not limited to test fixtures, markers, mocks, data generators, and custom assertions. For example, in SciPy, many APIs support input arrays of different backends, such as NumPy and PyTorch. To write backend-agnostic tests for these APIs, the project provides a fixture \code{xp}, and tests are expected to use \code{xp} to create arrays of different backends, instead of writing a test for each backend.

\subsection{Task 2: Test Integration}
The second subtask is to integrate the generated test case into the existing test suite of the project.
To this end, the approach should identify the appropriate location within the existing test suite where the test logically belongs, ensuring that it adheres to the conventions and utilities used in the existing tests:
\begin{definition}[Test integration task]
  Given a generated test case~$t$, identify the appropriate placement $(f_t, c_t, m_t)$ of $t$ within the existing test suite, where
  \begin{itemize}
    \item $f_t$ is the target test file in the test suite.
    \item $c_t::m_t$ are the target test class (or $\emptyset$ if no such class exists) and the target test method/function.
  \end{itemize}
\end{definition}
Besides identifying the correct placement location, the integration step should reuse existing test utilities. Generating a test case that increases patch coverage is insufficient to ensure that the test is actually integrated into the codebase.
Indeed, a survey of over 700 core project maintainers responsible for accepting contributions~\cite{gousiosWorkPracticesChallenges2015} reports that the most important factors influencing PR acceptance include \textit{code style} and \textit{technical fit}.
For example, this includes ``whether it adheres to the project conventions'', ensure ``keeping with the spirit of the project's other APIs'', and that ``its newly introduced code follows the total and functional style of the rest of the codebase''~\cite{gousiosWorkPracticesChallenges2015}.
Consequently, beyond merely increasing patch coverage, a generated test must also be \emph{well-integrated}, i.e., consistent with the style and conventions of the existing test suite and designed to reuse test utilities.

%% file: text_rev/approach.tex
\section{Approach}
\label{sec:approach}

\begin{figure*}[htbp]
  \centering

  \includegraphics[width=\linewidth]{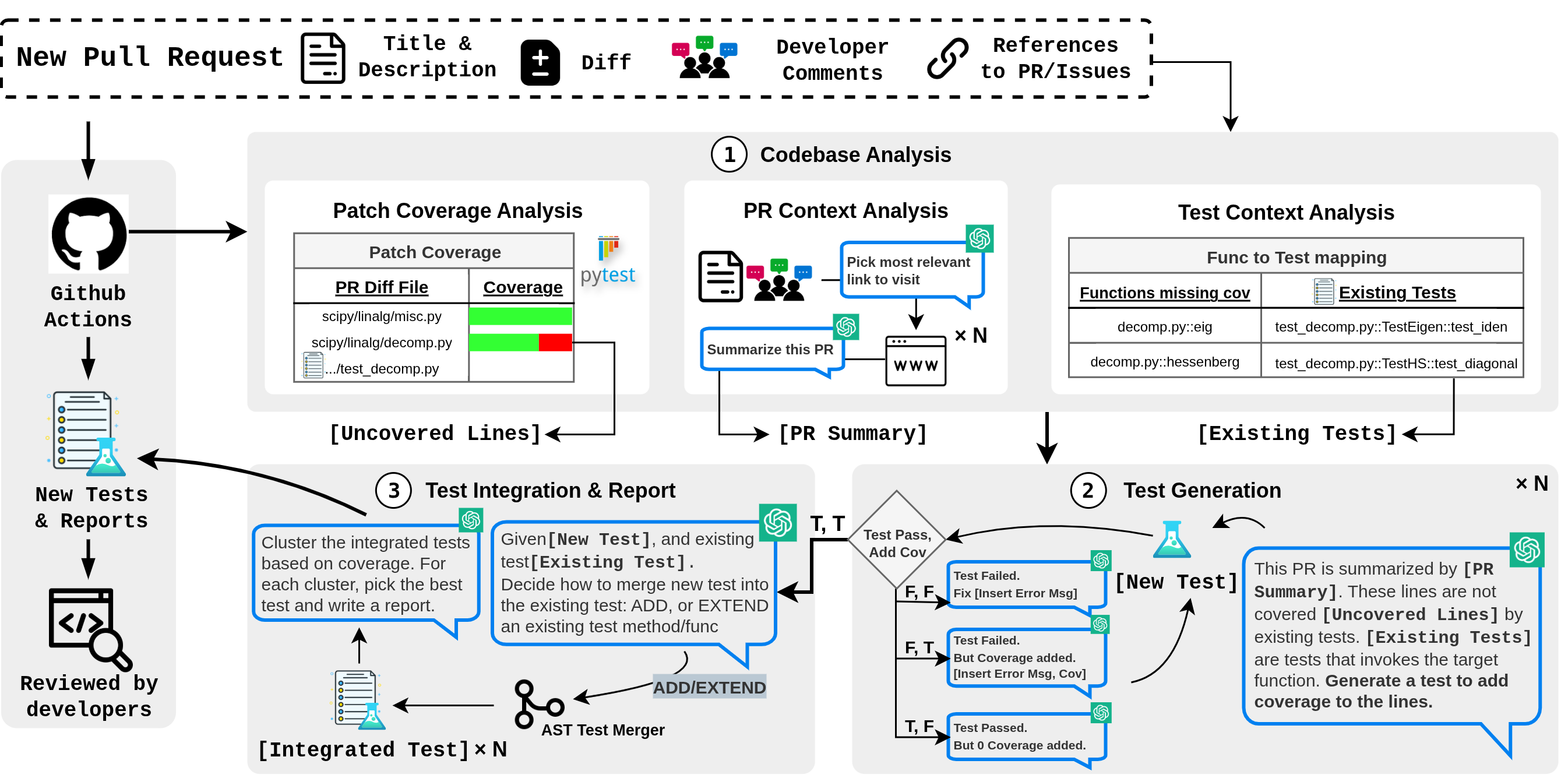}
  \caption{Flowchart of \methodName{} showing the main processing stages from PR input to integrated test case output.}
  \label{fig:flowchart}
\end{figure*}

Figure~\ref{fig:flowchart} illustrates our approach.
Given a PR, \methodName{} produces a test case that increases patch coverage and integrates it into the existing test suite.
The approach consists of three stages: (1) codebase analysis, (2) test generation, and (3) test integration.

The analysis stage collects information from three sources. First, patch coverage analysis computes the patch coverage of the PR diff, identifying uncovered lines that require additional testing (Section~\ref{sec:approach:patch_coverage}). If the PR's code changes are fully covered, \methodName{} terminates as there is no contribution to make.
Second, PR context analysis extracts information from the PR description and its linked resources, such as related issues and documentation (Section~\ref{sec:approach:pr_context_analysis}). %
By analyzing this PR context, generated tests should align with the PR's intent.
Third, test context analysis identifies existing tests to serve as examples and the context where the newly generated tests should be inserted into (Section~\ref{sec:approach:test_context}).

The test generation stage (Section~\ref{sec:approach:test_generation}) generates new test cases targeting uncovered lines. Each candidate test is executed against the post-PR code, and \methodName{} verifies whether it improves patch coverage. If the test fails or does not enhance coverage, \methodName{} iteratively refines the test using execution error messages as feedback, up to a predefined number of attempts. For successful tests, \methodName{} prompts the LLM to maximize coverage improvement.

Finally, the test integration and report stage (Section~\ref{sec:approach:test_integration}) incorporates successful (i.e., passing and coverage-improving) test cases into the existing test files. It reuses the most relevant location to insert the test into, as identified by test context analysis. \methodName{} employs an LLM to decide whether to add a new test method or extend an existing one, ensuring seamless integration into the project's testing conventions.
To allow developers to quickly assess the newly added test cases, \methodName{} also generates a report (Section~\ref{sec:approach:test_integration}) summarizing the added test cases, their purpose, and their impact on coverage.
If multiple tests add the same coverage, \methodName{} additionally prompts an LLM to select the best one for reporting, using three criteria: test worthiness, integration quality, and PR relevance.

From a developer's perspective, \methodName{} automatically analyzes PRs, generates relevant test cases, and produces a summary if it finds a test that increases coverage. We envision the approach to be used on open PRs, e.g., implemented as a GitHub action that runs on new PRs, providing a low-friction way to improve the test suite.

\subsection{Codebase Analysis}
\label{sec:approach:codebase_analysis}

The codebase analysis stage extracts contextual information that is useful for generating relevant test cases that are suitable for the existing codebase.
The approach performs three analyses, which are conceptually independent, as shown in Figure~\ref{fig:flowchart}.

\subsubsection{Patch Coverage Analysis}
\label{sec:approach:patch_coverage}

To understand the coverage gap of the PR, \methodName{} computes the patch coverage.
This analysis identifies uncovered lines within the PR diff by running the entire regression test suite.
If the patch coverage is 100\%, the process terminates, deeming the PR sufficiently tested.

To convey the missing coverage to the LLM, the approach annotates each uncovered line using the comment \texttt{\# UNCOVERED!}, ensuring the LLM focuses on these areas during subsequent stages.
The summary spans multiple files, concatenated for clarity. An example is provided in Figure~\ref{fig:uncovered_branches}, which highlights uncovered branches in the \texttt{status} method of \texttt{PrimitiveJob}.
\methodName{} segments the lines with missing patch coverage into \textit{focal functions}. It only considers lines that are inside functions or methods. This is because we observed that uncovered top-level lines often do not require dedicated tests, such as \code{if TYPE\_CHECKING: from scipy.\_lib import utils}.

We also acknowledge that not all uncovered lines necessarily merit additional tests. Prior work~\cite{Brandt2024CoverGaps,Sterk2024WhyIgnoreCoverageChecks} shows that sometimes coverage gaps are considered acceptable by developers when the code is unlikely to be buggy, is legacy and no longer expected to change, or is exercised by regular fuzz-testing runs.

\begin{figure}[t]
  \centering
\begin{lstlisting}[
      style=mypythonstyleDiff
    ]
# qiskit/primitives/primitive_job.py
class PrimitiveJob(BasePrimitiveJob[ResultT, JobStatus]):
...
  def __init__(self, function, *args, **kwargs):
...
  def status(self) -> JobStatus:
    if self._status is None:
      self._check_submitted()
      if self._future.running():
        @RED@return JobStatus.RUNNING  # UNCOVERED!@endRED@
      elif self._future.cancelled():
        @RED@self._status = JobStatus.CANCELLED  # UNCOVERED!@endRED@
      elif self._future.done() and self._future.exception() is None:
        self._status = JobStatus.DONE
      else:
        @RED@self._status = JobStatus.ERROR  # UNCOVERED!@endRED@
    return self._status
\end{lstlisting}
    \vspace{-.6em}
    \caption{Code where lines that miss test coverage are annotated with \code{\# UNCOVERED}.}
    \label{fig:uncovered_branches}
    \vspace{-2em}
\end{figure}

\subsubsection{Pull Request Context Analysis}
\label{sec:approach:pr_context_analysis}
This analysis extracts contextual information from the PR description, developer discussions, and automated infrastructure messages, summarizing it into a format suitable for LLM input during the test generation stage.
For example, a PR~$A$ may address a bug reported in issue~$B$, where issue~$B$ attributes the bug to changes introduced in PR~$C$. This creates a network of interconnected PRs or issues. De Souza et al.~\cite{desouzaRevealingSoftwareDevelopment2024} highlight the importance of the ``Is contextualized by'' link type, observed in 20.5\% of cases, which indicates that developers reference other PRs or issues to obtain configuration details, understand limitations, or find relevant examples.

Specifically, \methodName{} retrieves the PR page's HTML content, converts it to markdown, and extracts embedded links. The PR page includes developer comments, code reviews, and CI messages. By consolidating this information, \methodName{} provides rich context to inform subsequent stages, enabling targeted test generation.
\methodName{} enriches the initial summary by visiting selected links. \methodName{} employs an LLM-based link selection mechanism that prioritizes official documentation, community forums, and technical guides. \methodName{} applies heuristic filtering to exclude irrelevant links, such as GitHub navigation pages, providing the LLM with a curated list of meaningful links.
Each iteration selects an outgoing link from the PR (we set a maximum of three links), retrieves its content, and updates the summary. %

Summarization and updates are performed using a structured LLM prompt. The prompt includes the HTML content of the PR, the current summary, the selected link content (if any), asking the LLM to generate or update the PR summary.
\methodName{} tracks provenance throughout the enrichment process, recording visited URLs and LLM calls for transparency and reproducibility. This iterative approach enables \methodName{} to gather comprehensive information about the PR, enhancing the quality of the generated test cases. LLM prompts to select links and to summarize the content are available in the supplementary material.

\subsubsection{Test Context Analysis} \label{sec:approach:test_context}

\methodName{}'s tests should be consistent with the style of existing tests. Specifically:

\begin{enumerate}
  \item \emph{Test placement}: The new test should be placed in the correct test file, class, and be named appropriately.
  \item \emph{Test utilities}: If suitable, the new test should reuse existing test utilities, such as fixtures, markers, and helper functions.
  \item \emph{Test style}: The new test should follow the coding style of neighboring tests, such as use of type hints, trailing commas, and use of single vs.\ double quotes.
\end{enumerate}

To satisfy these requirements, it is crucial to learn from existing tests relevant to a focal function with missing coverage. We refer to these existing tests as the \textit{test context} of a focal function. %

\input{images/test_context_example_new}

We present a motivating example in Figure~\ref{fig:test_context_example}.
Suppose that a PR adds a new function for matrix decomposition as \code{scipy.linalg.mat\_decomp}.
Further suppose there are tests for matrix operations in the file \code{scipy/linalg/tests/test\_matrix.py}. Figure~\ref{fig:test_context} shows an example of the test context of the focal function. Test utilities include imports, top-level variable \code{dtypes}, the test class \code{TestMatrixOperations}, test fixture \code{setup\_method}, as well as test methods \code{test\_matrix\_det} and \code{test\_matrix\_rank}. Test style includes the naming convention: \code{test\_matrix\_\{func\_name\}}.

Figure~\ref{fig:test_context_example_good} shows a test generated using the test context, while Figure~\ref{fig:test_context_example_bad} shows one generated without it. A developer reviewing the test in Figure~\ref{fig:test_context_example_bad} would immediately spot several issues requiring manual fixes before the test can be merged. First, the test needs to be moved into the correct test class. Second, it fails to reuse existing test utilities, such as the fixture that sets the random seed. It also hardcodes test inputs instead of using parametrization for matrix shapes and data types, limiting its scope. Third, the assertion is a generic equality assertion instead of the preferred NumPy-specific assertion for matrix equality. Lastly, the test fails to check for a warning on empty matrices. In contrast, the test in Figure~\ref{fig:test_context_example_good} is well-placed, reuses test utilities and conforms to style, and can be accepted with minimal or no changes.

\input{code/test_context_algor_rev}

To extract test context of a focal function from the existing test suite, we employ a mixture of lightweight static analysis, dynamic analysis, and LLM-prompting.
Algorithm~\ref{alg:test_context_algor} shows the steps taken for extracting test context analysis, as explained in the following.

\paragraph{Static Test Context}
If the PR modifies test files, then \methodName{} considers them as the candidate test files (\textit{test\_files}$'$) that may contain the test context (line 1-3). Alternatively, if the PR does not modify any test file, we rank all test files in the project by the lexical similarity of their path (e.g., \code{scipy/linalg/tests/test\_matrix.py}) and paths of the source files modified by the PR (e.g., \code{scipy/linalg/matrix\_decomp.py}) (line 4-5). We find this simple technique to be very effective, because large projects usually have well organized test suites. After obtaining a shortlist of candidate test files, \textit{test\_files}$'$, we give the LLM the PR and a summary of each file, and let it confirm and pick the relevant test files (line 6).

\paragraph{Dynamic Test Context (line 8-12)}
The next step aims to identify relevant existing tests via dynamic analysis.
Once the test files (\textit{test\_files}) are identified, we run all of their tests with a profiler, constructing a dynamic call graph (line 8). If the dynamic call graph contains call chains where a test method invokes a focal function, we mark the test as a caller of the focal function (line 9-12).

\paragraph{LLM Test Context (line 13-17)}
When no test covers the focal function, dynamic call graph analysis cannot find its caller, and \methodName{} \textit{falls back to LLM-prompting} (line 13-16). We provide the LLM with the PR diff, a test file summary, and the focal function, asking it for the test class and the test method that is the most relevant to the focal function.

After the relevant test method is identified, we extract the top-level and class-level statements around the test, including any imports, fixtures, and other test utilities used by the caller test.
Collectively they constitute the test context of a focal method (lines~11 and~16). The test context is saved for use by the test generator.

On a high level, test context analysis serves a dual purpose. First, it informs the test generation stage by providing insights into existing tests, enabling the generation of new tests that mirror the style and taking advantage of proper test utilities. Second, it aids the test integration stage by identifying the most relevant test files and test classes for incorporating the generated tests seamlessly (Section~\ref{sec:approach:test_integration}). To optimize efficiency, \methodName{} maintains a cache of test context information, allowing it to reuse previously computed results and avoid redundant analysis in subsequent runs.

\subsection{Test Generation}
\label{sec:approach:test_generation}

Once the codebase analysis is complete, we generate test cases targeting the uncovered lines identified during patch coverage analysis. The test generation process is iterative and adaptive, leveraging the outputs of the codebase analysis stage: uncovered lines, PR context, and test context. These inputs are combined into structured prompts for the LLM. We detail the test generation step below.

The test generation conversation begins with a prompt that includes the PR's diff, the PR context, and the focal function annotated with uncovered lines (as shown in Figure~\ref{fig:uncovered_branches}). The prompt instructs the LLM to ``Inspect and summarize the lines modified by the PR that are uncovered by existing regression test suites.'' The LLM returns with a natural-language summary of why the lines are uncovered. Then, \methodName{} uses a prompt that includes the PR context, the summary of uncovered lines, and one test context (pick $t \sim TC[\textit{foc}]$, as provided by Algorithm~\ref{alg:test_context_algor}). The prompt instructs the LLM to ``generate test cases for the PR's changes'' and to return an executable \code{pytest} test (Figure~\ref{fig:test_context_example_good}).

In our preliminary experiments, we discovered that it is extremely rare for LLMs to generate a correct and useful test case on first trial. It is intuitive since human developers also need ``trial and error.'' In light of this and program repair approaches~\cite{xiaAutomatedProgramRepair2024, icse2025-RepairAgent, zhangAutoCodeRoverAutonomousProgram2024a}, \methodName{} employs a refinement technique to iteratively fix and improve generated tests. First, \methodName{} executes the test on the post-PR project to check whether it 1) passes and 2) covers previously uncovered lines. Refinement continues depending on the four different outcomes. If the test passes and adds new coverage, then the generated test is good and \methodName{} moves on to the test integration stage. If the test failed and added no coverage, \methodName{} prompts the LLM with the error message to fix the test. If the test failed but added coverage (which, based on our observations during the evaluation, often indicates that the test is correct except the oracle) \methodName{} prompts the LLM with the error message, the code with both added and missing coverage annotated with special comments, and ask it to fix the runtime error while preserving the coverage. If the test passed but does not add coverage, \methodName{} prompts the LLM to change the test to increase coverage. Three custom prompts contextualize the generated test with its runtime outcome, enabling the LLM to make informed improvements on the test.

\subsection{Test Integration and Report}
\label{sec:approach:test_integration}
\input{images/test_integration_example}

\methodName{}'s generated tests are standalone tests that need to be integrated into the test suite properly. With the help of test context, we can directly integrate a standalone test into its test context. The test context provides the existing test file and test class into which our generated test should be incorporated. At first, for each test and its context, \methodName{} prompts an LLM to decide on the integration mode: whether to add the test as a new test method/function or to extend the body of an existing test.
Then, the generated test, existing test file, and integration mode are passed to an abstract syntax tree (AST) transformer that performs the merge. The transformer checks if the generated test includes new imports, top-level variable definitions, test fixture definitions, and merges them into the existing test file accordingly. Figure~\ref{fig:test_integration_example} shows how the generated test in Figure~\ref{fig:test_context_example_good} is integrated into its test context. The test method \code{test\_matrix\_decompose} and the import of the focal method are inserted. Other imports and the helper method \code{setup\_method} are skipped because they are already present in the existing test file.

\methodName{} generates a report of its proposed test additions. For each PR, after generating $N$ tests, \methodName{} filters the successful (i.e., passing and coverage-improving) tests. In case of multiple successful tests, \methodName{} clusters them by the coverage addition. Each cluster contains tests that add the same set of line coverage; tests that add a strict subset of any cluster are dropped. 

While tests of a cluster provide identical coverage, they may still vary in quality. \methodName{} automates test selection. We prompt the LLM with the PR context, PR diff, PR's patch coverage, the tests in the cluster, and instruct it to select the highest-quality test. The quality criteria include: 1) \textit{worthiness}: how likely the test will catch regressions; 2) \textit{integration}: how seamlessly the test integrates with existing tests, considering style and use of test utilities; and 3) \textit{relevance}: how closely the test aligns with the PR's intention. We give hand-written positive and negative examples in the prompt to illustrate these criteria. Each selected test is then summarized using an LLM, which generates a concise report detailing the test, its purpose, and impact on coverage. At the end, \methodName{} attaches the report to the open PR for developer review.

%% file: images/test_context_example_new.tex
\begin{figure*}[t]
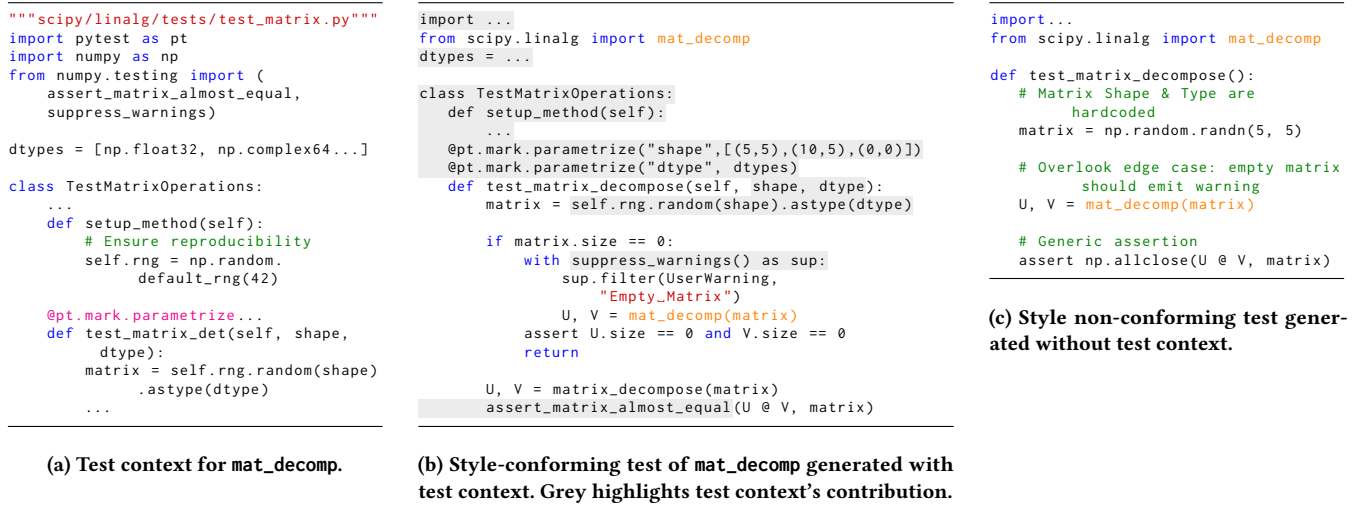

  \centering

  \begin{subfigure}[t]{0.28\linewidth}
    \centering
\begin{lstlisting}[
      style=mypythonstyle
    ]
"""scipy/linalg/tests/test_matrix.py"""
import pytest as pt
import numpy as np
from numpy.testing import (
    assert_matrix_almost_equal,
    suppress_warnings)

dtypes = [np.float32, np.complex64...]

class TestMatrixOperations:
    ...
    def setup_method(self):
        # Ensure reproducibility
        self.rng = np.random.default_rng(42)
    
    |@pt.mark.parametrize|...
    def test_matrix_det(self, shape, dtype):
        matrix = self.rng.random(shape).astype(dtype)
        ...
\end{lstlisting}
    \caption{Test context for \code{mat\_decomp}.}
    \label{fig:test_context}
  \end{subfigure}
  \hspace{1em}
  \begin{subfigure}[t]{0.40\linewidth}
    \centering
\begin{lstlisting}[
      style=mypythonstyleDiff
    ]
@HL@import ...@endHL@
from scipy.linalg import ^mat_decomp^
@HL@dtypes = ...@endHL@

@HL@class TestMatrixOperations:@endHL@
   @HL@def setup_method(self):@endHL@
@HL@       ...@endHL@
   @HL@@pt.mark.parametrize("shape",[(5,5),(10,5),(0,0)])@endHL@
   @HL@@pt.mark.parametrize("dtype", dtypes)@endHL@
   def test_matrix_decompose(self, @HL@shape, dtype@endHL@):
       matrix = @HL@self.rng.random(shape).astype(dtype)@endHL@

       if matrix.size == 0:
           with @HL@suppress_warnings() as sup:@endHL@
               sup.filter(UserWarning,
                   "Empty Matrix")
               U, V = ^mat_decomp(matrix)^
           assert U.size == 0 and V.size == 0
           return

       U, V = matrix_decompose(matrix)
       @HL@assert_matrix_almost_equal@endHL@(U @ V, matrix)
\end{lstlisting}
    \caption{Style-conforming test of \code{mat\_decomp} generated with test context. Grey highlights test context's contribution.}
    \label{fig:test_context_example_good}
  \end{subfigure}
  \hspace{1em}
  \begin{subfigure}[t]{0.265\linewidth}
    \centering
\begin{lstlisting}[
      style=mypythonstyleDiff
    ]
import...
from scipy.linalg import ^mat_decomp^

def test_matrix_decompose():
   # Matrix Shape & Type are hardcoded 
   matrix = np.random.randn(5, 5)

   # Overlook edge case: empty matrix should emit warning
   U, V = ^mat_decomp(matrix)^

   # Generic assertion
   assert np.allclose(U @ V, matrix)
\end{lstlisting}
    \caption{Style non-conforming test generated without test context.}
    \label{fig:test_context_example_bad}
  \end{subfigure}
\vspace{-.3em}

  \caption{Comparison of tests generated for
           \code{mat\_decomp} with and without providing the LLM with test context.}
  \label{fig:test_context_example}
\end{figure*}

%% file: code/test_context_algor_rev.tex
\begin{figure}
    \vspace{-1.6em}
\begin{minipage}[t]{0.49\textwidth}
\begin{algorithm}[H]
\caption{Extract Test Context} 
\label{alg:test_context_algor} 
\begin{algorithmic}[1]
\small
\Require $\delta, \mathcal{U}, T$            \Comment{PR code changes, Uncovered lines, All test-file paths}
\Ensure  $TC$                               \Comment{Mapping from focal functions to test context}

\For{$f \in \delta$}                         \Comment{\textbf{\textit{Static Test Context}}}
    \If{$f$ is a test file}
        \State $\textit{test\_files}'.\mathrm{append}(f)$
    \EndIf
\EndFor

\If{$\textit{test\_files}' = \emptyset$}      \Comment{no direct test files found}
    \State $\textit{test\_files}' \gets \textbf{Jaccard}(T,\ \delta)[1{:}K]$
\EndIf

\State $\textit{test\_files} \gets
       \textbf{LLM}(\textsc{PickTestFiles},\ \textit{test\_files}',\ \delta)$

\Statex\hrulefill

\State $TC \gets \emptyset$
\State \textit{trace} $\gets \mathrm{Profiler}(\textit{test\_files})$

\For{$\textit{foc} \in  \textbf{ParseFunc}(\mathcal{U})$}         \Comment{\textbf{\textit{Dynamic Test Context}}}
    \State \textit{callers} $\gets \textbf{AncestorsOf}(\textit{trace},\ \textit{foc})$
    \State $TC_{\textit{foc}} \gets
           [\textbf{Extract}(c) \mid c \in \textit{callers} \land c\ \text{is test function}]$
    \State $TC[\textit{foc}] \gets TC_{\textit{foc}}$
\EndFor

\For{$\textit{foc} \in  \textbf{ParseFunc}(\mathcal{U})$}           \Comment{\textbf{\textit{LLM Test Context (fallback)}}}
    \If{$TC[\textit{foc}] = \emptyset$}         \Comment{no test invoked the focal function}
        \State $t \gets
               \textbf{LLM}(\textsc{PickTestFunction},\ \textit{test\_files},\ \textit{foc},\ \delta)$
        \State $TC[\textit{foc}] \gets \textbf{Extract}(t, \textit{test\_files})$
    \EndIf
\EndFor

\State \Return $TC$
\end{algorithmic}
\end{algorithm}
\end{minipage}

\vspace{-1em}
\end{figure}

%% file: images/test_integration_example.tex
\begin{figure}[t]
    \centering
    \vspace{-.5em}
\begin{lstlisting}[
      style=mypythonstyleDiff
    ]
"""scipy/linalg/tests/test_matrix.py"""
@HL@import pytest as pt@endHL@
@HL@import numpy as np@endHL@
@HL@from numpy.testing import (@endHL@
@HL@    assert_matrix_almost_equal,@endHL@
@HL@    suppress_warnings,@endHL@
@HL@)@endHL@
+from scipy.linalg import mat_decomp

@HL@dtype_list = [np.float32, np.complex64, ...]@endHL@

@HL@class TestMatrixOperations:@endHL@

@HL@    def setup_method(self):@endHL@
@HL@        ...@endHL@

@HL@    def test_matrix_determinant(self):@endHL@
@HL@        ...@endHL@

@HL@    def test_matrix_rank(self):@endHL@
@HL@        ...@endHL@

+   @pt.mark.parametrize("shape",[(5, 5),(10, 5),(0, 0)])
+   @pt.mark.parametrize("dtype", dtype_list)
+   def test_matrix_decompose(self, shape, dtype):
+       matrix = self.rng.random(shape).astype(dtype)
+
+       if matrix.size == 0:
+           ...
+
+       U, V = mat_decomp(matrix)
+       assert...
\end{lstlisting}
    \caption{Integration of generated standalone test (shown in Figure~\ref{fig:test_context_example_good}) into \code{scipy/linalg/tests/test\_matrix.py}}
    \label{fig:test_integration_example}
    \vspace{-1em}
\end{figure}

%% file: text_rev/implementation.tex
\section{Implementation}
\methodName{} is implemented in Python and leverages containerization to ensure a consistent and isolated testing environment. For each project, a dedicated Docker container is created to execute the test suite as specified by documentation and continuous integration (CI) pipeline. This setup essentially simulates a CI pipeline locally.

To retrieve dynamic call chains and extract dynamic test contexts, we utilize viztracer~\cite{viztracerDoc} as the dynamic profiler. Additionally, \methodName{} employs DsPy~\cite{khattabDSPyCompilingDeclarative2023} to structure prompts. 
\methodName{}'s parameters include the choice of LLM, model-related parameters (e.g., temperature), \# test cases to generate per PR, and \# maximum feedback iterations---all exposed via a configuration file. \methodName{} passes the model-related parameters to DsPy, which abstracts and handles the communication with the LLM. 
In total, \methodName{} uses eleven structured prompts to achieve its objectives. Figure~\ref{fig:flowchart} illustrates eight of these prompts for simplicity. The other three prompts not shown are used for 1) picking the most relevant test file, 2) picking relevant test class and test method, and 3) summarizing the uncovered lines. We refer readers to \methodName{}'s repository for the complete set of prompts.

%% file: text_rev/eval.tex
\section{Evaluation}
We address the following research questions:
\begin{enumerate}
    \item \textbf{RQ1: Effectiveness}: How effective is \methodName{} at producing tests that \textit{pass} and that \textit{add patch coverage}?
    \item \textbf{RQ2: Acceptability}: To what extent are the tests generated by \methodName{} acceptable by developers in terms of their added value for detecting future bugs, their integration into the existing test suite, and their relevance to the PR?
    \item \textbf{RQ3: Ablation Study}:
        \begin{enumerate}
            \item \textbf{Test context component}: How do the two kinds of test context (dynamic and LLM-based) contribute to \methodName{}'s effectiveness?
            \item \textbf{Feedback component}: What is the impact of \methodName{}'s use of runtime feedback on test pass rate and coverage?
        \end{enumerate}
\end{enumerate}

\input{tables/prselection}

\paragraph{Project Selection}
We evaluate our approach on three open-source projects: SciPy, Qiskit, and Pandas.
We selected them as they are:
1) complex and production-quality, necessitating thorough testing, 2) highly active, with many PRs and contributors, 3) already having a comprehensive test suite, which matches our ``last-mile'' approach of adding new tests to PRs.
SciPy and Pandas are popular libraries for scientific computing and data science. Qiskit is the most popular compiler framework for quantum computing, which helps us understand \methodName{}'s effectiveness on newer domains.
Compared to prior work on automated test generation for Python, which are evaluated on smaller projects~\cite{pizzornoCoverUpEffectiveHigh2025, xieChatUniTestChatGPTbasedAutomated2023, fraserEvoSuiteAutomaticTest2011}, our selection is more challenging as it evaluates the augmentation of already strong test suites of complex codebases.

\paragraph{PR Selection}

We systematically filter PRs to identify suitable ones for evaluation. We begin with the most recent 2,000 PRs in June 2025, and apply the following filters: 1) the PR is merged, 2) it contains code changes other than deletions and documentation changes, 3) the PR title must not contain specific keywords, such as "DOC" or "backport", and the PR must not modify files outside our coverage tracking scope (e.g., we ignore Pandas PRs that modify files in the \code{pandas/io} directory as they are responsible for (de)serialization to various formats), and 4) the PR modifies no more than five code files, which helps us filter out large refactorings. Finally, we run the regression test suite on the PR's branch, and retain the PRs whose patch coverage is not 100\%. The final selection of PRs is shown in Table~\ref{tab:prselection}.

To conduct experiments and balance the projects, we randomly sample 50 PRs of Qiskit and Scipy from Table~\ref{tab:prselection}, plus all 45 PRs of Pandas, resulting in a total of 145 PRs to evaluate RQ1. The qualitative analysis and test submission (RQ2) and the ablation study (RQ3) are done on a smaller benchmark of 30 PRs, where we downsample to 10 PRs per project.

\paragraph{Model and Parameters}
For our evaluation, we use GPT-4o-mini~\cite{gpt4ominirelease} for its cost-effectiveness in large-scale test generation. Using GPT-4o-mini also minimizes data contamination, as its training data predates all the 145 PRs used in our evaluation. 
We configure the model with a temperature of 0.7, which we empirically found to balance test diversity and quality. \methodName{} exposes the choice of LLM and parameters via a configuration file, allowing easy parameter tuning and experimentation with different LLMs. 

\paragraph{Metrics} These are the metrics used for evaluation.
\begin{enumerate}
  \item \textbf{Test pass rate}: The ratio of tests that pass, divided by the total number of tests generated.
  \item \textbf{Coverage increment}: The number of newly covered lines that generated tests add to the project.
  \item \textbf{Test review scores}: In RQ2.1, we manually review tests and score them on worthiness, integration, and relevance.
  \item \textbf{Test submission outcomes}: In RQ2.2, we track the number of tests submitted as PRs and their outcomes.
  \item \textbf{Cost}: The monetary cost (USD) of using the LLM.
\end{enumerate}

\subsection{RQ1: Effectiveness}
\input{tables/effectiveness}
\input{tables/coverage.tex}

We apply \methodName{} on the 145 PRs and measure three key metrics: test pass rate, coverage increment, and cost.
Table \ref{tab:pass_rate} shows the test pass rate of the tests generated by \methodName{} on the 145 benchmark PRs. Table \ref{tab:coverage} shows the coverage increment. \methodName{} fully covers \totalPRFullCov{} PRs with 100\% patch coverage. Importantly, each line of coverage added by \methodName{} \textit{represents a readable, regression-quality test}, and should not be compared with coverage obtained from approaches, such as fuzzing or search-based software testing (SBST). We demonstrate in RQ2 that these tests are acceptable by project maintainers with high success rate.
Studying the costs of running \methodName{}, we find that augmenting one PR costs between \$0.08 and \$0.17, depending on the project, with an overall average of \avgCostPerPR{} per PR.
We expect this cost to decrease as LLMs become cheaper in the future.

\begin{resultbox}
  \methodName{} reliably generates passing tests that add high-quality regression test coverage (\totalCovLinesAdded{} lines) to complex, real-world projects, showing the benefit of integrating LLM-based test generation into CI workflows. A substantial portion (\ratioPRFullCov{}) of PRs reach full patch coverage, highlighting the value of targeted, PR-context-aware test generation. The cost per PR remains affordable (\avgCostPerPR{} per PR), supporting the feasibility of large-scale adoption. Overall, these findings confirm the effectiveness of \methodName{} in bridging coverage gaps.
\end{resultbox}

\subsection{RQ2: Acceptability}

We envision \methodName{} to be incorporated into Continuous Integration (CI) pipelines, where it can automatically monitor and fix missing patch coverage in new PRs.
We ran the full \methodName{} pipeline to generate tests for each PR with missing test coverage. Then we performed two complementary experiments: (1) qualitative analysis, where reviewers score each generated test on worthiness, integration, and relevance; and (2) test submission, where tests are submitted to the upstream projects and developer feedback is analyzed.

Unlike RQ1, which focuses on quantitative metrics, RQ2 delivers a qualitative and fine-grained assessment of \methodName{}. We also present a case study to illustrate how \methodName{} helps increase test coverage, exposes issues, and motivates feature changes.

\subsubsection{RQ2.1: Qualitative Analysis}
\label{sec:eval:qualitative_analysis}

We evaluate 10 PRs per project, selected from the 50 PRs used in RQ1. Two reviewers independently assess each test with three criteria: (1) \emph{Worthiness}: the test's potential to detect regressions or bugs (e.g., tests for trivial getters are less valuable); (2) \emph{Integration}: the test's conformity to the structure of the existing test suite, including its placement in files/classes and adherence to project-specific conventions, such as test utilities; and (3) \emph{Relevance}: the test's alignment with the intent of the original PR. Each criterion is rated on a scale from 1 (poor) to 5 (good). Developer feedback (Section~\ref{sec:eval:test_submission}) confirmed that these three criteria are instrumental for accepting tests.

The two reviewers both have eight years of experience in software engineering. Reviewer A is a contributor to quantum computing platforms, including Qiskit, PennyLane, BQSKit and PyTket. Reviewer B has contributed tests to SciPy, Pandas, and is also experienced in software testing. To standardize the review process, we provide a step-by-step guideline that defines the task, the three scoring criteria (with examples). For example, the guideline instructs the reviewers to first review the PR itself and understand why patch coverage is incomplete; then the reviewers examine the test report generated by \methodName{}, which includes a summary, coverage visualization, runtime log, test patch, and the complete test file. 

We acknowledge that this qualitative assessment is inherently subjective. For instance, determining the "most appropriate" test utilities for the integration score depends on deep, project-specific knowledge that even experienced developers may not possess for all submodules. Therefore, the reviewers provided scores on a best-effort basis, reflecting their expert but not infallible judgment.

\begin{figure}[t]
  \centering
  \vspace{-1em}
  \includegraphics[width=0.98\linewidth]{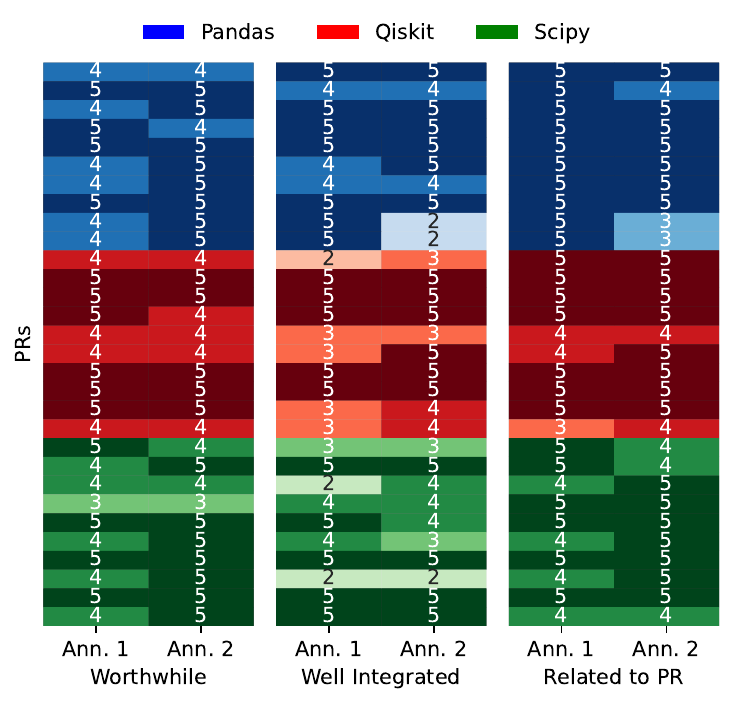}
  \vspace{-1em}
  \caption{Heatmap of annotation scores provided by reviewers for each PR.}
  \label{fig:heatmap_annotations}
  \vspace{-1em}
\end{figure}

Figure~\ref{fig:heatmap_annotations} presents a heatmap summarizing the annotation scores assigned by reviewers for each PR. The color coding indicates the project, while the degree of shading represents the score (darker is better).
We report a satisfactory level of agreement between reviewers accounting for \percAgrWorthwhile{} for worthiness, \percAgrWellIntegrated{} for integration, and \percAgrRelatedToPr{} for relevance. The average scores across all PRs are \avgScoreWorthwhile{} for worthiness, \avgScoreWellIntegrated{} for integration, and \avgScoreRelatedToPr{} for relevance.

The worthiness score of \avgScoreWorthwhile{} indicates that the tests were generally effective in addressing coverage gaps and preventing regression. Reviewers noted their utility in covering untested paths, branches, or exceptional cases, with comments like ``nice to cover an extra branch.'' However, feedback also highlighted areas for improvement, such as weak or absent assertions, with critiques like ``no oracle'' or ``\code{assertIsNotNone} is weak,'' suggesting that while some tests added coverage, their bug detection abilities could be further improved.

Integration scored \avgScoreWellIntegrated{}, suggesting that tests generally adhered to the suite's structure and conventions, which supports maintainability. Comments like ``Good location'' or ``right class'' confirmed test placement. A common suggestion is to use better test parameterization. Inconsistent use of frameworks, such as \code{unittest} over \code{pytest}, was also noted.

Relevance scored highest at \avgScoreRelatedToPr{}, indicating strong alignment with PR intent.
This confirms our design decision to use PR context, such as the PR description and discussion, to guide test generation.

\subsubsection{RQ2.2: Test Submission as Pull Requests} \label{sec:eval:test_submission}
\begin{figure}
    \centering
    \includegraphics[width=0.98\linewidth]{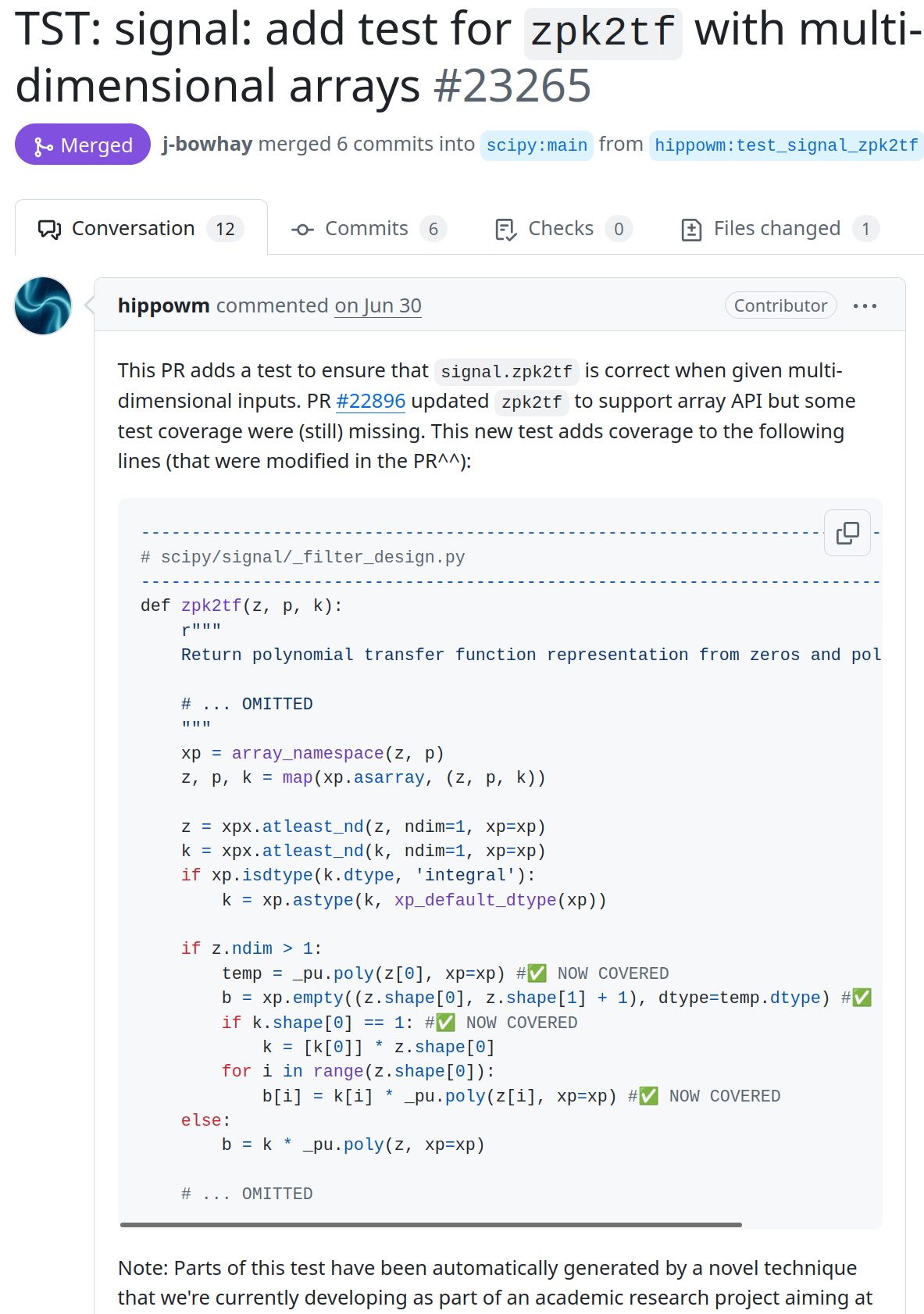}
    \caption{Example Test Submission \href{https://github.com/scipy/scipy/pull/23265}{(PR 23265)} to SciPy. This test uncovers a bug in the previously-untested branch of \code{signal.zpk2tf} when input array \code{z} is multi-dimensional. The test was accepted and merged after fixing the CI failures.}
    \vspace{-1.6em}
    \label{fig:test_submission_example}
\end{figure}

Our north-star goal is to submit these tests as PRs fully automatically. However, submitting a large volume of PRs with generated tests is impractical. Therefore, we submitted a subset of tests to obtain feedback from developers. %
Each PR submission includes the following. An example submission is shown in Figure~\ref{fig:test_submission_example}.

\begin{itemize}
  \item A manually written title and description.
  \item A reference to the original PR with missing coverage.
  \item A visualization of coverage addition.
  \item A note on AI usage and human review before submission.
\end{itemize}

We submitted a total of \totalSubmittedTests\, \methodName{}'s tests (9 for SciPy, 1 for Pandas, 2 for Qiskit) in a best-effort manner. All 12 tests received some developer feedback, 8 have been accepted and merged, 2 are open, 2 have been rejected.

Developers appreciated the increased test coverage. Interestingly, our submissions prompted discussion about AI usage for test generation; most developers were supportive, though some raised concerns about maintainability. We addressed these by highlighting our manual review before submission. Most tests did not raise concerns from developers regarding their worthiness (i.e., potential to catch future bugs) or relevance to the original PR. Two tests were rejected due to concern of worthiness. Developers commented that "this particular test is not very strong," and "not sure it's super valuable."
Notably, 3/\totalSubmittedTests~tests received developer suggestion to \textit{use more appropriate test utilities and follow specific conventions}, the largest challenge, despite \methodName{}'s use of test context analysis. 
This is consistent with RQ2.1 (Section~\ref{sec:eval:qualitative_analysis}), where integration scored the lowest. Specifically, the three suggestions are 1) adding a custom test marker to skip on certain backends (Section~\ref{sec:eval:case_study}), 2) switching to an assertion customized for checking array equality, and 3) using a custom context manager for setting options.

The selection of test utilities is highly nuanced and context-dependent. Taking custom assertions as an example, there are too many in projects such as SciPy and Pandas. Choosing the appropriate one depends on several factors: the functionality under test (e.g., exact versus approximate matching), the types of arguments involved (e.g., integer vs. floating-point), whether or not FP exceptional values should be handled (e.g., \code{[NaN] == [NaN]}?), whether or not comparisons should be type-flexible (e.g., \code{[1] == [1.0]}?). This complexity means that LLMs often need to reason and adapt test utilities from the test context, rather than simply replicating.

The test utilities are distributed across multiple namespaces and locations, making retrieval challenging. For example, when adding a new test method, such as \code{scipy/module/test\_something.py::TestSomething::test\_method\_new}, relevant test utilities may be defined in various places: global libraries like \code{pytest} and \code{numpy.testing}, project-specific modules such as \code{scipy.\_lib}, configuration files like \code{scipy/conftest.py}, submodule-level utilities (e.g., \code{scipy.signal}), as well as file-level and class-level utilities. This dispersion means that gathering all possible test utilities for consideration is non-trivial, and the correct utility may be missed or overlooked by automated approaches.

\vspace{-.3em}
\begin{resultbox}
  \methodName{}'s tests are generally well-received by developers (8/12 already merged). The tests score well on worthiness and relevance, though integration remains a challenge. Developer discussions highlight both enthusiasm for AI-assisted testing and concerns about maintainability, which we address through transparency and collaboration. Overall, these findings demonstrate the acceptability of \methodName{}'s tests in real-world development workflows.
\end{resultbox}
\vspace{-.35em}

\subsubsection{Case Study of Test Submission} \label{sec:eval:case_study}
We present a case study to show how \methodName{} can help increase project coverage, uncover issues in the codebase, and motivate feature changes.
Figure~\ref{fig:test_submission_example} is a screenshot of the test submitted to SciPy. This test ensures the focal function \code{signal.zpk2tf} is correct when input array \code{z} is multi-dimensional. The entire branch \code{if z.ndim > 1} is untested, in which the original PR modified four lines.

The submitted test case passed locally (because we only submit passing, coverage-adding tests), but failed on two checks in Scipy's CI. This can happen because \methodName{}'s existing implementation is limited to a docker-based build of the project, while SciPy's CI builds the project across multiple library dependencies, operating systems, and architectures. The test fails 1) on GPUs, and 2) when array backend is JAX \cite{jax} or \code{array-api-strict} \cite{arrayapistrict}. The test failed on GPUs as expected because the focal function \code{signal.zpk2tf} does not support running on GPUs as of 1.16. To solve this, as instructed by the developers, we manually added a marker \code{@skip\_xp\_backends(cpu\_only=True)} to the test.

The other CI failure uncovers a bug in \code{signal.zpk2tf}. It is expected to work on backends JAX and \code{array-api-strict}. However, \methodName{}'s test exposed that the previously-untested branch fails on these two backends. This led the developers to fix the implementation of \code{signal.zpk2tf}. In the discussion, one developer proposed a patch that fixes the bug on \code{array-api-strict} backend. After the CI failures were resolved by adding test markers, the test PR was approved and merged into SciPy as part of its 1.17.0 milestone. The bug-fix patch was also submitted and merged. %

\subsection{RQ3: Ablation Study}
\input{tables/ablation_rev}

We perform an ablation study to evaluate the contributions of \methodName{}'s two key components: (Dynamic) Test Context and Feedback Component. To do this, we created the following variants of \methodName{}: \textbf{\methodName{} (LLM Test Context)} uses exclusively LLM to generate test context (which is the fallback mode of \methodName{} when no DTC is available). \textbf{\methodName{} (No Test Context)} does not perform test context analysis and removed test context in all prompts. \textbf{\methodName{} (No Feedback)} removes test generation stage's feedback loop that provides the LLM with runtime error and coverage information. By default, \methodName{} uses a maximum of three rounds of feedback.

Table~\ref{tab:ablation} shows the results on 30 PRs (10 per project). Overall, \methodName{} outperforms all ablated variants in terms of coverage increment, in all three projects. In particular, \methodName{} using dynamic test context outperforms two variants (i) purely LLM-generated test context, (ii) no text context, by the same \avgCovImprvOverLLMPercent{} in coverage increment, highlighting the importance of precise retrieval of test context thanks to dynamic analysis. \methodName{} (No Feedback) performs the worst in terms of both metrics, underscoring the importance of iterative test improvement using runtime feedback. In terms of pass rate, \methodName{} (No Test Context) performs the best in Qiskit and Pandas, but adding fewer lines of coverage. This suggests that while test context helps generate tests that add more coverage, it may also introduce complexity that can lead to more test failures.

\begin{resultbox}
  The use of (dynamic) test context significantly improves the chance of increasing coverage by \avgCovImprvOverLLMPercent{}, compared to using (i) no test context and (ii) LLM-generated test context. The latter can be very imprecise that the performance is on par with not using test context at all. Iterative feedback significantly improves the test pass rate by \avgPassRateImprvOverNoFBPercent{} and coverage by \avgCovImprvOverNoFBPercent{}.
\end{resultbox}

%% file: tables/prselection.tex
\begin{table}[t]
\caption{Projects and PRs used in evaluation.}
\vspace{-.1em}
\label{tab:prselection}
    \centering
\begin{adjustbox}{width=\linewidth}
\setlength{\tabcolsep}{3pt}
\begin{tabular}{@{}lccccc>{\bfseries} c@{}}
\toprule
Project & \multicolumn{6}{c}{Pull requests} \\
\cmidrule{2-7}
& Considered & Merged & \makecell{With code\\ changes} & \makecell{Keyword \&\\ scope}  & $\leq$ 5 Files & \makecell{Incomplete\\ patch cov.}\\
\midrule
SciPy & 2,000 & 1,822 & 851  & 772 & 572 & 121\\
Qiskit & 2,000 & 1,842 & 919  & 712 & 481 & 135 \\
Pandas & 2,000 & 1,763 & 813 & 372 & 273 & 45\\
\bottomrule
\end{tabular}
\end{adjustbox}
\vspace{-1em}
\end{table}

%% file: tables/effectiveness.tex
\begin{table}[t]
\caption{\textbf{Test pass rate} and \textbf{cost}.}
\vspace{-.2em}
\label{tab:pass_rate}
    \centering
\begin{adjustbox}{width=0.45\textwidth}
\begin{tabular}{lcc >{\bfseries}c>{\bfseries}c}
\toprule
Project & Tests Generated & Passed & Pass Rate & Cost (\$ per PR)\\
\midrule
SciPy & 832  & 274 & 32.9\% & 0.17\\
Qiskit & 689 & 170 & 24.7\% & 0.08 \\
Pandas & 737 & 161 & 21.8\% & 0.08\\
\bottomrule
\end{tabular}
\end{adjustbox}
\vspace{-.7em}
\end{table}

%% file: tables/coverage.tex
\begin{table}[t]
    \caption{\textbf{Coverage increment}. PRs Considered is the total number of PRs with non-full patch coverage used in this experiment. Cov Added is the number of PRs that \methodName{} successfully added coverage to. $\rightarrow$ 100\% Patch Cov is the number of PRs that reached full patch coverage after the test addition. \#lines added is the sum of line coverage added by \methodName{} on all PRs. Ratio is the average ratio of coverage added / coverage missing across PRs.}
    \vspace{-.2em}
\label{tab:coverage}
    \centering
\begin{adjustbox}{width=0.45\textwidth}
\begin{tabular}{lcc >{\bfseries} ccc}
\toprule
Project & PRs Considered & Cov Added & $\rightarrow$ 100\% Patch Cov &\#lines added & \% Ratio \\
\midrule
SciPy & 50 & 27 & 15 & 112 & 40.0\% \\
Qiskit & 50 & 21 & 13 & 57 & 34.0\% \\
Pandas & 45 & 17 & 14 & 20 & 33.5\%\\
\bottomrule
\end{tabular}
\end{adjustbox}
\vspace{-.2em}
\end{table}

%% file: tables/ablation_rev.tex
\begin{table*}[t]
  \caption{Performance of \methodName{} and its ablated variants. 
  }
  \vspace{-.8em}
  \label{tab:ablation}
  \centering
  \small
  \newcolumntype{?}{!{\vrule width 1pt}}
  \begin{tabular}{>{\bfseries} l?*{3}{ccc}}
    \toprule
    \multirow{2}{*}{Method} &
    \multicolumn{3}{c}{Qiskit} &
    \multicolumn{3}{c}{Scipy}  &
    \multicolumn{3}{c}{Pandas} \\
    \cmidrule(lr){2-4}\cmidrule(lr){5-7}\cmidrule(lr){8-10}
      & \textbf{Pass Rate} & \textbf{\# Lines} & \textbf{Cost(\$)} &
        \textbf{Pass Rate} & \textbf{\# Lines} & \textbf{Cost(\$)} &
        \textbf{Pass Rate} & \textbf{\# Lines} & \textbf{Cost(\$)} \\
    \midrule
    \methodName{}                             & 28.8\%  & \textbf{30}  & 0.15    & \textbf{38.7\%}   & \textbf{18} & 0.21    & 19.2\%           & \textbf{8}  & 0.09  \\
    \midrule
    \methodName{} (LLM Test Context)        &  29.5\%       &  19   & 0.14    & 28.8\%            & 2           & 0.15    & 18.8\%           & 7           & 0.09  \\
    \methodName{} (No Test Context)         &    \textbf{31.5\%}    &  21   & 0.14    & 31.3\%            & 2           & 0.19    & \textbf{27.7\%}  & 5           & 0.09  \\
    \midrule
    \methodName{} (No Feedback)             &     7.6\%     &  5   & 0.04    & 12.4\%            & 2           & 0.06    & 3.5\%            & 3           & 0.03  \\
    \bottomrule
  \end{tabular}
  \vspace{-.5em}
\end{table*}

%% file: text_rev/threats.tex
\vspace{-.5em}
\section{Threats to Validity}
\paragraph{Construct Validity} RQ2.1 relies on two reviewers acting as proxies for the original developers. These reviewers may lack the deep contextual knowledge of the project and the specific intent behind a code change that the original developers possess. Consequently, their judgment on the quality (worthiness, integration, and relevance) of a generated test might differ from that of the code's author. Moreover, these quality metrics are subjective, and standards vary among developers (e.g., how strong a test must be to catch regressions), making consistent measurement difficult. These factors may affect the validity of the findings for RQ2.1.

\paragraph{Internal Validity}
Our approach relies on automated setup for test execution environments across multiple projects. 
as changes to projects' build systems may affect the reproducibility of our tool's execution. 
For instance, we observed that SciPy periodically updates its build process, necessitating updates to our Docker configurations.
Moreover, the effectiveness of the LLM-based test generation depends on the prompt engineering and the specific model used, introducing potential variability in the measured test quality. Additionally, the absence of guardrails to prevent LLM hallucinations could result in the generation of irrelevant or incorrect tests. 
Also, \methodName{} assumes that the PR is bug-free, leading it to generate only passing tests. In practice, its utility could be enhanced by integrating it with other automated testing techniques that first identify and filter out buggy pull requests. 

\paragraph{External Validity} 
Our evaluation focused on only three open-source Python projects (SciPy, Qiskit, Pandas), which may not represent the full spectrum of software projects. 
Their codebases were likely part of the LLMs’ training data, so the effectiveness of \methodName{} on projects not seen during training needs further investigation. 
The generalizability of our results to other languages (e.g., C++, Java) or project domains remains uncertain. 
Furthermore, development practices regarding test coverage vary across projects, e.g., sometimes coverage gaps are ignored or intentional~\cite{Brandt2024CoverGaps,Sterk2024WhyIgnoreCoverageChecks}, which may influence the utility of our tool.

%% file: text_rev/related_work.tex
\vspace{-.5em}
\section{Related Work}

\paragraph{Automated Test Generation Techniques}
Traditional test generation techniques remain relevant alongside LLM-based approaches. Search-based software testing (SBST) employs metaheuristic algorithms to optimize test creation for specific goals like coverage. Tools like EvoSuite~\cite{fraserEvoSuiteAutomaticTest2011} and Pynguin~\cite{lukasczykPynguinAutomatedUnit2022}  generate tests by exploring code structure and exercising diverse execution paths. Similarly, DSpot~\cite{Danglot2019DSpot} aims to enhance the bug-finding ability of test suites by applying custom mutations to existing tests and using mutation score as a fitness function. This line of work, in principle, can be adapted to improve patch coverage.
Automated test generation spans diverse methodologies, including specification-based approaches~\cite{boyapatiKoratAutomatedTesting2002}, feedback-directed random testing~\cite{pachecoFeedbackDirectedRandomTest2007}, and symbolic execution-guided techniques~\cite{godefroidDARTDirectedAutomated2005}. Transformer-based methods~\cite{tufanoUnitTestCase2021} leverage machine learning advancements to enhance test generation. 
In contrast, we propose a PR-specific test generation approach that uses PR context to improve patch coverage.

\paragraph{Test Generation with Large Language Models}
LLMs have reshaped automated test generation~\cite{Lemieux2023,Nan2025TestIntention,pizzornoCoverUpEffectiveHigh2025,Ryan2024CodeAwarePromptingAS,schaferEmpiricalEvaluationUsing2024,icse2026-Testora,Meta2024LLMUTGen,Hossain2025LLMOracleTOGLL,wangHITSHighcoverageLLMbased2024,Yuan2024ImproveChatGPTUTGen,Deljouyi2025UTGen}. CoverUp~\cite{pizzornoCoverUpEffectiveHigh2025} uses LLMs for coverage-guided test generation, targeting uncovered code regions. SymPrompt~\cite{Ryan2024CodeAwarePromptingAS} introduces code-aware prompting strategies, decomposing test generation into multi-stage sequences aligned with execution paths.
TestPilot~\cite{schaferEmpiricalEvaluationUsing2024} generates tests using function signatures and documentation, eliminating the need for additional training data.
UTGen~\cite{Deljouyi2025UTGen} integrates an LLM into SBST to enhance the understandability of generated tests.
Unlike these works, we integrate dynamic analysis with LLM-based techniques to extract and utilize test context, addressing integration challenges overlooked by prior methods.

\paragraph{Pull Request Testing}
EvoSuiteR~\cite{shamshiriAutomatedUnitTest2015} generates regression tests by comparing two versions of a Java class, while Testora~\cite{icse2026-Testora} uses LLMs to detect unintended behavioral changes in PRs. These approaches focus on regression or behavioral testing rather than patch coverage. 
CTG~\cite{camposContinuousTestGeneration2014} combines EvoSuite with CI, focusing tests on on class complexity and project evolution. In contrast, we focus on generating tests for uncovered lines in PRs to fix patch coverage.

\paragraph{Developer Perceptions of Generated Tests}
Recent studies~\cite{Danglot2019DSpot, Brandt2024contributionstudy} investigated developer perceptions of automatically generated/augmented tests. Brandt et al.~\cite{Brandt2024contributionstudy} submitted pull requests to 39 Java projects with one DSpot-augmented~\cite{Danglot2019DSpot} test each, 19 PRs were accepted. 
The main findings were that manual edits are often necessary for the tests to be accepted; common edits include aligning assertion styles, relocating tests, and removing unnecessary code. The first two edits are addressed by \methodName{}'s use of test context. UTGen~\cite{Deljouyi2025UTGen} carried out a controlled experiment with 32 developers and shows that LLM-improved test understandability leads to better bug-fixing from developer feedback.